\documentclass{eas}
\usepackage{graphicx}
\usepackage{natbib}

\runningtitle{Juri\'c \& Ivezi\'c: SDSS, LSST and Gaia}

\TitreGlobal{Gaia: At the Frontiers of Astrometry}
\idline{EAS Publications Series, Volume 45, 2010, 281}
\Editors{Editors: C. Turon, F. Meynadier and F. Arenou.}
\begin{document}

\title{SDSS, LSST and Gaia: Lessons and Synergies}
\author{Mario Juri\'c}\address{Hubble Fellow; Harvard-Smithsonian Center for Astrophysics}
\author{\v{Z}eljko Ivezi\'c}\address{University of Washington and University of Zagreb}
\begin{abstract}
  
The advent of deep, wide, accurate, digital photometric surveys
exemplified by the Sloan Digital Sky Survey (SDSS) has had
a profound impact on studies of the Milky Way.
In the past decade, we have transitioned from a scarcity to an (over)abundance of precise,
well calibrated, observations of stars over a large fraction of the Galaxy. 
The avalanche of data will continue throughout this 
decade, culminating with Gaia and LSST.
This new reality will necessitate changes in methodology,
habits, and expectations both on the side of the large survey projects 
as well as the astrophysics community at large. We argue, based on 
the experience with SDSS, that surveys should release data as early and often 
as possible incorporating incremental improvements in each 
subsequent release, as opposed to holding off for a single,
big, final release. The scientific community will need to reciprocate 
by performing analyses and (re-analyses) appropriate to the 
current fidelity of the released data, understanding that these are 
continually evolving and improving products.

\end{abstract}
\maketitle

\section{ Introduction }

\begin{figure*}[!t]
\includegraphics[width=1.0\hsize,clip]{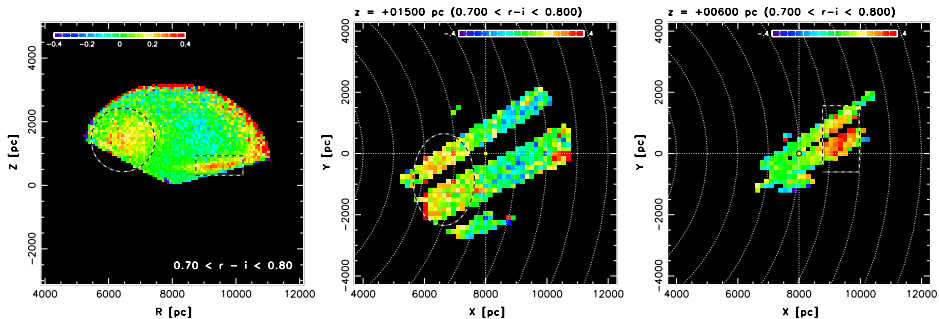}
\caption{
Two low-contrast disk overdensities detected in SDSS DR3 stellar 
number density map residuals. The left panel shows the $Data/Model-1$
residuals in the $(R, Z)$ plane, while the center and left panels
show the $X-Y$ cross sections intersecting the detected overdensities.
See \cite{J08} for details.
}
\label{clumps}
\end{figure*}

In the past decade, the data from the Sloan Digital Sky Survey (SDSS) have drammatically 
enhanced our picture of the Milky Way.
This includes comprehensive Galactic structure studies \citep[e.g.][]{carollo07, J08, I08, B10},
discoveries of overdensities and tidal streams \citep[e.g.][]{newberg02,belokurov06,J08} as 
well as discoveries of a significant population of ultra faint Milky Way satellites 
\citep[e.g.][and others]{willman05,belokurov07}. An example from \cite{J08}
given in Figure~\ref{clumps} shows a detection of two low-contrast disk overdensities in
stellar number density maps constructed from SDSS observations of $\sim48$ million stars.

The SDSS is a digital photometric and spectroscopic survey.
It has covered a contiguous region of about 7,600 deg$^{2}$ centered on the North 
Galactic cap, a smaller but deeper area in the Galactic South ($\sim225$ deg$^{2}$), and 
approximately $3,200$deg$^{2}$ of imaging for the Sloan Extension for Galactic 
Understanding and Exploration (SEGUE). More generally, it is an example of the kind of deep, wide and 
accurate surveys that will be increasingly common in this decade. A non-exhaustive
list includes Pan-STARRS PS1 \citep{kaiser02} that has already surpassed the 
SDSS in terms of covered area in less than 6 months
after the start of its science mission, SkyMapper (Keller et al. 2005), currently
in commissioning, the Large Sky Area Multi-Object Fibre Spectroscopic Telescope 
\citep[LAMOST;][]{su98}, the Dark Energy Survey \citep[DES;][]{des}, 
and of course Gaia \citep{gaia} and LSST \citep{tyson02,lsst}.

All these, including Gaia, share the common thread of aiming to produce and
publish large ($>10^8$ objects) and wide area datasets, a challenge the 
SDSS has already faced and successfully tackled.

\section{ Publishing and Consuming Large Datasets: Early, Often, Iterate }

The lessons learned from SDSS are many. Here we limit ourselves to only 
two specific areas of immediate interest to 
Gaia and its users: deciding when and what to publish, and appropriately 
approaching the published data.

\subsection{ Release Early, Release Often }

The SDSS has published the first public release of its data (the ``Early Data
Release''; EDR) in June 2001 \citep{stoughton02}. It consisted of roughly $460$ 
deg$^2$ of imaging data, $54,000$ follow-up spectra (about 5\% of the 
planned totals) and a photometric and astrometric catalog of  $\sim$14 million objects.

By today's SDSS Collaboration standards, the data included in this release would 
be considered substandard, even embarrassing.
To illustrate this point, at the time of the EDR even the true filter bandpasses 
were not known, the photometric calibration was at 3-5\% level (as opposed to
current $\sim 1-2$\%), and the output
formats and tables in the data distribution system were far from being
finalized. 

In spite of these deficiencies, the EDR was a major success (and never criticized
as premature). First, it demonstrated
that the survey was collecting and processing data (an important milestone
for a project 10 years in the making). Secondly, the release generated invaluable
feedback from the community which was incorporated into, and significantly strenghtened 
the subsequent data releases. Most importantly, the data described in the 
EDR produced substantial scientific returns, including (of importance to
Galactic astronomy) the discovery of the Monoceros Stream \citep{newberg02}.
At the time of writing of this contribution, the EDR paper has been cited 1215 times.

The SDSS has continued with releases on a 12-18 month schedule, with 
the latest (8$^{\rm th}$) data release planned for December 2010.
Besides adding more area, every new release involves
reprocessing of {\it all previously published data} to correct problems 
identified in the older releases, as well as to benefit from major improvements 
in the processing software \citep[for example, the inclusion of an improved photometric
calibration algorithm;][]{padmanabhan08}. In SDSS experience, many of the problems
reported by users were very subtle and practically the only way to discover them was to
perform ``cutting-edge" science analysis and then critically examine the results. 

The SDSS is far from alone 
in this approach. The RAVE survey \citep{rave}, as well as UKIDSS \citep{ukidss}, follow 
the same strategy. The planned Large Synoptic Survey Telescope (LSST) 
will follow a similar yearly release schedule.
This ``release early, release often'' approach is also well known in
open source software development \citep{raymond97}. The early data releases result in better
communication and feedback from the wider astrophysical community,
improve the quality of the published datasets, and reduce
the overall ``time to science''. 

We feel the same model is applicable 
to Gaia. In addition to helping 
the Gaia project team discover and correct problems as they appear, early 
releases will surely generate a significant amount of follow-up
science, enable synergies with ground based surveys such as LSST
(see Section~\ref{LSST}), as well as benefit the project in 
terms of education and public outreach.

\subsection{ Understanding that Datasets Evolve }

The paradigm described above puts an important onus of 
responsibility on the users to take into account the evolving 
nature of the datasets.
As will be noted elsewhere in this volume (see the contribution by 
Hogg), the only ``final'' product a survey can ever deliver are 
the {\em raw images}. Everything else, including the catalogs, is a 
{\em derived product} and subject to change as the understanding of the instrument,
the processing algorithms, or the underlying assumptions (priors) evolve.
To the best of our knowledge, no survey ever had all of these requirements well known early.  
However, the benefits of the early and regular access to data as they
are collected by far outweigh difficulties associated with evolving
datasets.

The end users need to be aware of these caveats, especially when making use of
the early data releases that will not have the full 
quality and reliability traditionally expected from a published
catalog. Of course, while certainly a potential source of frustration, good 
understanding of (any!) dataset and its limits
significantly improves the quality and longevity of the 
resulting science. Again, early and frequent data releases not only help
users to understand ``the survey error bars", but perhaps more importantly
help the project team to improve them before it is too late to ``fix the problem".

\section{ The Galaxy with LSST and Gaia }
\label{LSST}

\begin{figure}
\centering\includegraphics[width=0.75\linewidth]{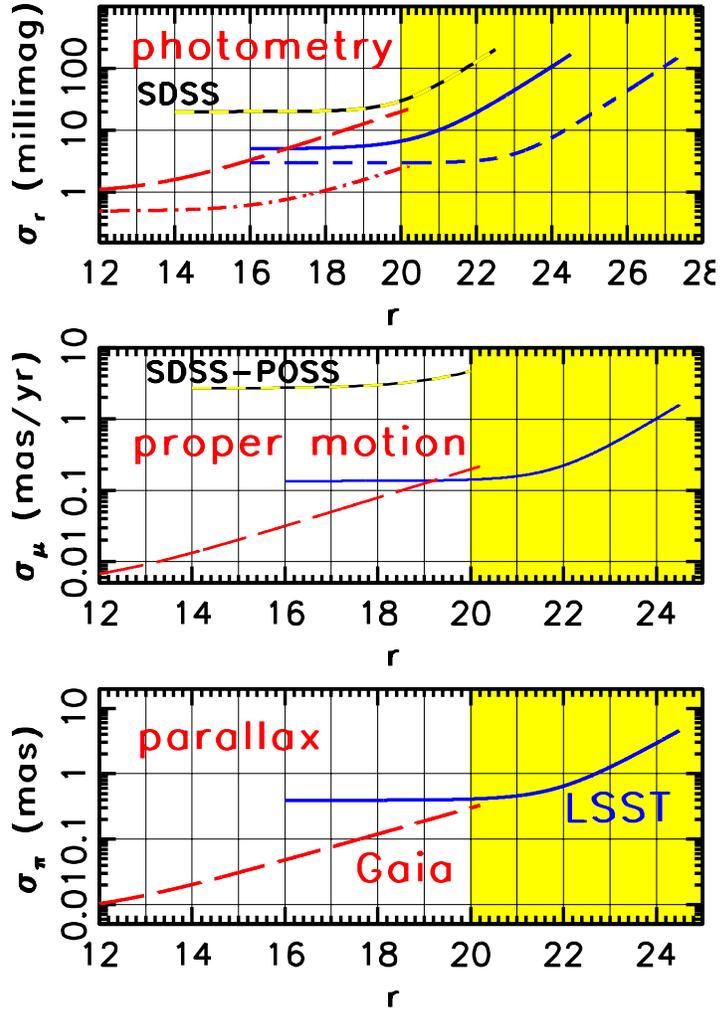}
\caption{A comparison of photometric, proper motion and parallax errors
for SDSS, Gaia and LSST, as a function of apparent magnitude $r$, for  
a G2V star (it is assumed that $r=G$, where $G$ is the Gaia's broad-band
magnitude). In the top panel, the curve marked ``SDSS'' corresponds
to a single SDSS observation. The red curves correspond to Gaia; the
long-dashed curve shows a single {\it transit} accuracy, and the    
dot-dashed curve the end of mission accuracy (assuming 70 transits).
The blue curves correspond to LSST; the solid curve shows a single  
{\it visit} accuracy, and the short-dashed curve shows accuracy for 
co-added data (assuming 230 visits in the $r$ band). The curve marked
``SDSS-POSS'' in the middle panel shows accuracy delivered by the
proper motion catalog of \citet{munn04}. In the middle and bottom
panels, the long-dashed curves correspond to Gaia, and the solid curves
to LSST. Note that LSST will smoothly extend Gaia's error vs.~magnitude
curves four magnitudes fainter. The assumptions used in 
these computations are described in Eyer et al. (in prep.).}
\label{LSSTvsGaia}
\end{figure}

The Large Synoptic Survey Telescope, \cite[LSST;][]{lsst}, will be a 
large, wide-field ground-based system designed to obtain 
multiple images covering the sky visible from its location at 
Cerro Pach\'{o}n, Chile. The current baseline design, with an 8.4m (6.7m effective) 
primary mirror, a 9.6 deg$^2$ field of view, and a 3,200 Megapixel camera, 
will allow about 10,000 square degrees of sky to be covered using pairs of 
15-second exposures in two photometric bands every three nights on average. 
The survey area will include 
30,000 deg$^2$ with $\delta<+34.5^\circ$, and will be imaged multiple times 
in six bands, $ugrizy$, covering the wavelength range 320--1050 nm. About 
90\% of the observing time will be devoted to a deep-wide-fast survey mode 
which will observe a 20,000 deg$^2$ region about 1000 times in six bands 
during anticipated 10 years of operations. These data will result in 
databases including about 20 billion objects. 

LSST will produce a massive and exquisitely accurate photometric and astrometric dataset
for about 10 billion Milky Way stars. The coverage of the Galactic plane
will yield data for numerous star-forming regions, and the $y$ band data 
will penetrate through the interstellar dust layer. Photometric metallicity
measurements  will be available for about 200 million main-sequence F/G stars 
which will sample the halo to distances of 100 kpc \citep{I08}. No other existing or 
planned survey will provide such a massive and powerful dataset to study the 
outer halo (including Gaia which is flux limited at $r=20$, and Pan-STARRS
which will not have the $u$ band). The LSST in its standard surveying mode
will be able to detect RR Lyrae and classical novae out to 400 kpc, 
and hence explore the extent and structure of the halo 
out to half the distance to M31. All together, the LSST will enable 
studies of the stellar distribution beyond the presumed edge of the 
Galactic halo, of their metallicity distribution 
throughout most of the halo, and of their kinematics beyond the thick
disk/halo boundary \citep{scibook}.

In the context of Gaia, the LSST can be thought of as its deep complement.
A comparison of LSST and Gaia performance is given in Figure~\ref{LSSTvsGaia}.
Gaia will provide an all-sky catalog with unsurpassed trigonometric parallax,
proper motion and photometric measurements to $r\sim20$ for about $10^9$
stars. LSST will extend this map to $r\sim27$ over half of the sky, detecting
about $10^{10}$ stars. Because of Gaia's superb astrometric and photometric quality,
and LSST's significantly deeper reach, the two surveys are highly complementary:
Gaia will map the Milky Way's disk with unprecedented detail, and LSST will extend
this map all the way to the halo edge (Eyer et al., in prep).

\acknowledgements 

MJ wishes to acknowledge the support for this work provided by NASA through 
Hubble Fellowship grant HF-51255.01-A awarded by the Space Telescope Science 
Institute, which is operated by the Association of Universities for 
Research in Astronomy, Inc., for NASA, under contract NAS 5-26555.
\v{Z}I  acknowledges support by NSF grant AST 07-07901, and by NSF grant 
AST 05-51161 to LSST for design and development activity.



\begin{thebibliography}{99}

\bibitem[Belokurov et al.(2006)]{belokurov06} Belokurov, V., et al. 2006, ApJL, 642, L137
\bibitem[Belokurov et al.(2007)]{belokurov07} Belokurov, V., et al. 2007, ApJ, 654, 897
\bibitem[Bond et al.(2010)]{B10} Bond, N.~A., et al.\ 2010, ApJ, 716, 1 
\bibitem[Carollo et al.(2007)]{carollo07} Carollo, D., et al.\ 2007, Nature, 450, 1020
\bibitem[Flaugher, et al.(2005)]{des} Flaugher, B. 2005, International Journal of Modern Physics A, 20, 3121
\bibitem[Ivezi\'{c} et al.(2008a)]{I08} Ivezi\'{c}, \v {Z}., Sesar, B., Juri\'{c}, M., et al.\ 2008a, ApJ, 684, 287 (I08)
\bibitem[Ivezi\'{c} et al.(2008b)]{lsst} Ivezi\'{c}, \v {Z}., Tyson, J.A., Allsman, R, et al.\ 2008b, arXiv:0805.2366
\bibitem[Juri\'{c} et al.(2008)]{J08} Juri\'{c}, M., Ivezi\'{c}, \v{Z}., Brooks, A., et al.\ 2008, ApJ, 673, 864 (J08)
\bibitem[Kaiser et al.(2002)]{kaiser02} Kaiser, N., et al.\  2002, Proceedings of the SPIE, 4836, 154 
\bibitem[Lawrence et al.(2007)]{ukidss} Lawrence, A., et al.\  2007, MNRAS, 379, 1599 
\bibitem[LSST Science Collaborations et al.(2009)]{scibook} LSST Science Collaborations, et al.\ 2009, ``LSST Science Book, v2.0'', arXiv:0912.0201 
\bibitem[Munn et al.(2004)]{munn04} Munn, J.~A., et al.\ 2004, AJ, 127, 3034
\bibitem[Newberg et al.(2002)]{newberg02} Newberg H.J., Yanny, B., Rockosi, C., et al. 2002, ApJ, 569, 245
\bibitem[Padmanabhan et al.(2008)]{padmanabhan08} Padmanabhan, N., et al.\ 2008, ApJ, 674, 1217 
\bibitem[Perryman et al.(2001)]{gaia} Perryman, M.~A.~C., et al.\ 2001, A\&A, 369, 339 
\bibitem[Raymond(1997)]{raymond97} Raymond, E. S., ``The Cathedral and the Bazaar'', http://www.catb.org/\~{}esr/writings/cathedral-bazaar/cathedral-bazaar/ar01s04.html
\bibitem[Stoughton et al.(2002)]{stoughton02} Stoughton, C., Lupton, R.H., Bernardi, M., et al. 2002, AJ, 123, 485
\bibitem[Su et al.(1998)]{su98} Su, D.~Q., Cui, X., Wang, Y., \& Yao, Z.\ 1998, Proc. SPIE, 3352, 76 
\bibitem[Tyson(2002)]{tyson02} Tyson, J.~A.\ 2002, Proceedings of the SPIE, 4836, 10 
\bibitem[Willman et al.(2005)]{willman05} Willman, B., et al.\ 2005, ApJL, 626, L85 
\bibitem[Zwitter et al.(2008)]{rave} Zwitter, T., et al.\ 2008, AJ, 136, 421

\end{thebibliography}
\end{document}